\documentstyle[12pt]{article}
\def\lapp{{\ \lower 0.6ex \hbox{$\buildrel<\over\sim$}\ }}
\def\gapp{{\ \lower 0.6ex \hbox{$\buildrel>\over\sim$}\ }}
\def\beq{\begin{equation}}
\def\eeq{\end{equation}}
\def\beqn{\begin{eqnarray}}
\def\eeqn{\end{eqnarray}}
\def\to{\rightarrow}

\def\qq{q\bar{q}}

\def\pp{{\rm p\bar{p}}}

\def\qbar{{\bar{q}}}
\def\bbar{{\bar{b}}}
\def\bb{{b\bar{b}}}
\def\tbar{{\bar{t}}}
\def\tt{t\bar{t}}
\def\ww{W^+W^-}

\def\degree{^{\circ}}

\def\GeV{{\rm GeV}}

\def\TeV{{\rm TeV}}

\def\cF{{\cal F}}

\def\gt{\Gamma_t}
\def\iN{{1\over N}}
\def\iiN{{2\over N}}
\def\akikii{\widehat{k_1k_2}}
\def\aqiqii{\widehat{q_1q_2}}
\def\akiqii{\widehat{k_1q_2}}
\def\akiqi{\widehat{k_1q_1}}
\def\akiiqi{\widehat{k_2q_1}}
\def\akiiqii{\widehat{k_2q_2}}
\def\aqiqi{\widehat{q_1q_1}}
\def\aqiiqii{\widehat{q_2q_2}}
\def\akipi{\widehat{k_1p_1}}
\def\aqipi{\widehat{q_1p_1}}
\def\akipii{\widehat{k_1p_2}}
\def\aqipii{\widehat{q_1p_2}}
\def\akiipi{\widehat{k_2p_1}}
\def\aqiipi{\widehat{q_2p_1}}
\def\akiipii{\widehat{k_2p_2}}
\def\aqiipii{\widehat{q_2p_2}}
\def\apipii{\widehat{p_1p_2}}
\def\apiipii{\widehat{p_2p_2}}
\def\apipi{\widehat{p_1p_1}}

\def\cF{{\cal F}}

\def\kinem{1}
\def\threed{2}
\def\slicesa{3}
\def\slicesb{4}
\def\ptet{5}
\def\etapd{6}
\def\models{7}
\def\cosbg{8}
\def\asym{9}

\begin{document}
\begin{titlepage}
\vspace*{-1cm}
\begin{flushright}
DTP/94/60   \\
UR-1365 \\
ER-40685-815 \\
July 1994 \\
\end{flushright}
\vskip 1.cm
\begin{center}
{\Large\bf
Additional Soft Jets in ${\rm t\bar{t}}$ Production \\[3mm]
at the Tevatron $\pp$ Collider}
\vskip 1.cm
{\large Lynne H. Orr}
\vskip .2cm
{\it Department of Physics, University of Rochester \\
Rochester, NY 14627-0171, USA }\\
\vskip   .4cm
and
\vskip .4cm
{\large  W.J. Stirling}
\vskip .2cm
{\it Departments of Physics and Mathematical Sciences, University of Durham \\
Durham DH1 3LE, England }\\
\vskip 1cm
\end{center}
\begin{abstract}
A large fraction of top quark events in $\pp$ collisions
at $1.8\ \TeV$ will contain additional soft hadronic jets from
gluon bremsstrahlung off the quarks and gluons  in the hard
processes $\qq, gg \to \tt \to \bb \ww$. These extra jets
can cause complications when attempting to reconstruct $m_t$
from the invariant mass of combinations of final-state quarks and
leptons.  We show how such soft radiation cannot be
unambiguously associated with either initial-state radiation or
or with final-state radiation off the $b$ quarks. The top quarks
can radiate too, and in fact the pattern  of radiation has a very rich
structure, which depends on the orientation of the final-state
particles with respect to each other and with respect to the beam.
We calculate the full radiation pattern of soft jets in the soft
gluon approximation and compare with several approximate forms
which are characteristic of parton shower Monte Carlos.  The
implications for top mass measurements are discussed.
\end{abstract}
\vfill
\end{titlepage}
\newpage
\section{Introduction}

Very recently the CDF collaboration has reported evidence
for a top quark of mass $174 \pm 17\  \GeV/c^2$ in $\pp$ collisions
at $1.8\ \TeV$ \cite{CDFTOP}. The leading-order production processes are
$\qq, gg \to \tt \to \bb \ww$, and a statistically significant signal
is seen in the channels $\ww\to l\nu l \nu$ and $\ww\to l \nu \;
+ \; \mbox{jets}$ ($l=e,\mu$). From the latter sample,
the top mass is reconstructed from the
final-state lepton and jet momenta. A potential problem with this procedure
arises when the final state contains an additional hadronic jet, since
it is not known {\it a priori} whether this jet arises
from initial-state radiation, in which case it should be ignored,
or whether it originated (say) in bremsstrahlung off a $b$ quark, in
which case it should be included in the $m_t$ reconstruction. We note that a
substantial fraction of the candidate top events reported by CDF do indeed
have an additional jet over and above the number associated
with the leading-order production and decay process.

In fact
there is {\it no} meaningful separation of such jets into
`initial-' and `final-state' radiation \cite{KOS}. In any part
of phase space, the soft jet cross section
contains contributions from radiation off the incoming $\qq, gg$,  off the
produced $\tt$ at the production  stage, off the $t $
and  $b$ quarks in the $t\to bW$  decay process,
and from interferences between these.
In specific regions, however, we might expect certain types of radiation
to be more important than others. For example, we would expect
 radiation close
to the beam axis to  be dominated by  gluon emission off the incoming partons,
while radiation close to a $b$ quark should be  dominated by collinear
gluon emission off that quark. But away from these special collinear
regions, all particles (including the top quarks) emit  gluons
with approximately equal strength.
In what follows, we will
study the distribution of soft gluon radiation over  all of phase space,
to see to what extent our expectations about initial- and final-state radiation
 are valid.  We will see that while the initial-/final-state radiation
picture is too na\"{\i}ve,
a decomposition of the radiation pattern into contributions associated with
$\tt$ production, decay, and their interference provides the
information necessary for top mass reconstruction.

Our purpose in this paper is twofold.  First, we wish to study the
correct distribu\-tions\footnote{Our calculations are
`exact' in the sense that we include gluon radiation
 off all colored particles in the production and decay process. Our
 only approximation is to assume that the gluon is `soft', i.e.
 $E_g/E \ll 1$, where $E \sim m_t$
  is a typical energy scale of the subprocess.} to determine
where the gluons come from and where they go, in a way that is directly
relevant to top momentum reconstruction and mass measurement.
Our second aim is
to compare these results with simpler models
which are characteristic  of the way gluon  radiation is
 implemented in Monte Carlo event generators, such as the initial-/final-state
picture mentioned above.
The difference between these approximations and the exact
 treatment could then provide an indication of the errors implicit in using
 certain Monte Carlos to generate extra soft jets in $\tt$ production.


\section{General formalism for soft gluon radiation}

\subsection{Gluon radiation in $\tt$ production and decay}

We start with the leading-order process
\begin{eqnarray}
a(k_1) + b(k_2) \to  t(q_1) + \bar{t}(q_2) \to
b(p_1) + W^+ + \bar{b}(p_2) + W^- \>,
\end{eqnarray}
where $ab = q\bar q$ or $gg$, and the particles' momenta are indicated in
parentheses. Naively, the invariant
mass of each $b-W$ system is equal to $m_t$. Now suppose that the final state
contains an extra
 gluon, with momentum $k^\mu$.
In terms of Feynman diagrams, this gluon can be emitted off
(taking the $q\bar q$ process as an example) either of the incoming
light quarks, the $s$-channel gluon, off either of the top quarks
 before they decay weakly ({\it i.e.}, on the timescale of the strong
 production process), off either of the top quarks
on the timescale of their  weak decay, or off
either of the final-state $b$ quarks (and also
from the decay products of a hadronically decaying $W$). All of these
amplitudes can of course interfere.

Because of the infra-red divergence associated with
gluon emission, an extra jet in a $\tt$ event will usually
be soft. In this case, one can analyze the radiation pattern
in the soft gluon approximation, as was done in Refs.~\cite{KOS,KOSHAD}.
One can write
\begin{equation}
{1\over d\sigma_0}\ {d\sigma\over d E_g\> d\cos\theta_g \> d\phi_g}\ = \
{\alpha_s\over 4 \pi^2} \ E_g \ \cF \; ,
\label{softsigma}
\end{equation}
where $d\sigma_0$ is the differential cross section for the lowest-order
process ({\it i.e.}, with no gluon radiation), $E_g$ is the energy of the
soft gluon, and $\alpha_s$ is the strong  coupling.
The function $\cF$ is the sum of \lq antenna patterns' of the
radiation from the different sources listed above.
It can be written generically as
\begin{equation}
\cF = \cF_{\mbox{\tiny PROD}} + \cF_{\mbox{\tiny DEC}} +\cF_{\mbox{\tiny INT}}
\end{equation}
where $\cF_{\mbox{\tiny PROD}}$ is the contribution from
emission at the $\tt$ production stage (including initial-state radiation),
$\cF_{\mbox{\tiny DEC}}$ is the contribution from emission
off the $t$ and $\bar t$
and their decay products at  the weak decay stage,
and $\cF_{\mbox{\tiny INT}}$ refers to the
interferences between these emissions.
More specifically, $\cF_{\mbox{\tiny INT}}$ contains contributions from
the interference (i) between radiation in the $t$ decay  and
that in the $\tbar$ decay and (ii) between radiation in
$\tt$ production and radiation in either decay.\footnote{
In terms of the results of Ref.~\cite{KOS},
$\cF_{\mbox{\tiny PROD}}$ corresponds to $|A|^2$,
$\cF_{\mbox{\tiny DEC}}$  to $|B_1|^2 + |B_2|^2$,
and $\cF_{\mbox{\tiny INT}}$  to
$- 2 {\rm Re} [B_1 B_2^*]+ 2 {\rm Re}[A(B_2 - B_1)^*]$.}
Note that for soft gluons with $E_g \sim \Gamma_t \sim 1\ \GeV$,
$\cF_{\mbox{\tiny INT}}$ is sensitive to the top decay width, as discussed
in Ref.~\cite{KOS}.
However, the observable soft jets that are relevant to the $\pp$ collider
experiments have energies much larger than $\Gamma_t$, and so in practice
the interference terms are numerically small.

Explicit expressions for  $\cF$ for the $\qq$ and $gg$ subprocesses have been
presented in \cite{KOSHAD} and are listed again here for completeness:
\begin{eqnarray}
\cF{\mbox{\tiny PROD}}  &=& c_1 \akikii + c_2 \akiqi +
c_3 \akiqii + c_4 \akiiqi
+ c_5 \akiiqii + c_6 \aqiqii + c_7 \aqiqi + c_8 \aqiiqii\; , \nonumber \\
\cF{\mbox{\tiny DEC}}  &=& c_7 [\aqiqi + \apipi -2\aqipi ]
+ c_8 [\aqiiqii+ \apiipii -2\aqiipii ]\; , \nonumber \\
\cF{\mbox{\tiny INT}}&=&
 \chi_1\; \left\{ c_2 [ \akipi -\akiqi ] +c_4 [ \akiipi -\akiiqi ]
 +c_6 [ \aqiipi -\aqiqii ] + 2  c_7 [ \aqipi -\aqiqi ]  \right\} \nonumber \\
 &+& \chi_2\; \left\{  c_3 [ \akipii -\akiqii ] +c_5 [ \akiipii -\akiiqii ]
 + c_6 [ \aqipii -\aqiqii ] +2  c_8 [ \aqiipii -\aqiiqii ] \right\}
  \nonumber \\
 &+& \chi_{12} \; c_6 [ \apipii -\aqipii  -\aqiipi + \aqiqii ] \; ,
\label{general}
\end{eqnarray}
where the antennae $\widehat{pq}$  and `profile functions' $\chi_i$
are defined by
\begin{eqnarray}
\widehat{pq} \> & = & \> {p\cdot q \over p\cdot k \ q\cdot k} \; ,\nonumber \\
\chi_i & = & {m_t^2\Gamma_t^2 \over (q_i\cdot k)^2 + m_t^2 \Gamma_t^2 }
\qquad (i=1,2)\; , \nonumber \\
\chi_{12} & = & {m_t^2\Gamma_t^2\; (q_1\cdot k\; q_2\cdot k + m_t^2\Gamma_t^2)
 \over \left[ (q_1\cdot k)^2 + m_t^2 \Gamma_t^2 \right]
    \; \left[ (q_2\cdot k)^2 + m_t^2 \Gamma_t^2 \right] } \; .
\label{chidef}
\end{eqnarray}
In terms of the gluon energy, we have $\cF \sim E_g^{-2}$, and so
the cross section has the infra-red
behavior  $d \sigma / d E_g \sim E_g^{-1}$, as expected.
Additional collinear
singularities arise from the $p\cdot k$ denominators when $p^2 = 0$.

The antenna coefficients $c_i$ depend on the color structure of the
hard scattering and are different for the $\qq$ and $gg$ processes
($C_F = 4/3,\ N=3$):
\begin{center}
\begin{tabular}{|c|c|c|}  \hline
\rule[-1.2ex]{0mm}{4ex}  coefficient & $\qq\to\tt$ & $gg\to\tt$ \\ \hline
 $c_1$       &  $-\iN$   &   $-2C_F+2N+2Y$    \\
 $c_2$       &  $2C_F-\iN$   & $C_F-X-Y$     \\
 $c_3$       & $\iiN$    &  $C_F+X-Y$    \\
 $c_4$       & $\iiN$    &  $C_F+X-Y$    \\
 $c_5$       & $2C_F-\iN$    &    $C_F-X-Y$    \\
 $c_6$       & $-\iN$    &  2Y     \\
 $c_7$       & $-C_F$    &  $-C_F$    \\
 $c_8$       &  $-C_F$   &  $-C_F$    \\
  \hline
\end{tabular}
\end{center}
The quantities $X$ and $Y$ depend on the $gg\to\tt$ subprocess energy
and scattering angle. Explicit expressions and a discussion can be found in
\cite{KOSHAD}.

In general terms, the antennae which contribute to  $\cF{\mbox{\tiny PROD}}$
correspond to color strings stretched between  the initial-state
partons and the final-state $t$ quarks {\it before} they decay, and those
which contribute to  $\cF{\mbox{\tiny DEC}}$ correspond
to  color strings linking the
 final-state $t(\bar t)$ and $b(\bar b)$ quarks at the weak decay stage.
We remind the reader that color antennae exhibit the `string effect':
radiation between the two momenta in an antenna is enhanced compared to that
outside them; see for example Ref.~\cite{BOOK}.
We will see several instances below where this effect
influences gluon distributions.

For purposes of top momentum reconstruction and mass measurement, this
decomposition serves as a guide to what to do with additional soft jets from
gluons.  Gluons from the production stage are emitted before the $t$ and
$\tbar$ quarks go on shell and should not be included in top momentum
reconstruction.  Those from the decay stage {\it are} to be included along with
the top decay products.  For the interference contributions no such
clear--cut assignment is possible; however for the examples we consider here,
the interference contribution is negligible.

\subsection{Gluon radiation in simpler models}

There are several simpler models which we could construct if we were
interested in simulating gluon radiation in top events without using the
full distribution, Eq.~(\ref{general}).
Consider, for example, a model which includes
only initial-state radiation. This would be appropriate, for example, for
the purely electroweak process $q \bar q \to Z^0 \to l^+l^-$, where
the color string is stretched between the incoming quarks.
In the language of the antennae introduced above, this corresponds to
\begin{equation}
\cF \; =\;  2 C_F  (2N) \;   \akikii  \; ,
\label{isronly}
\end{equation}
for $\qq (gg)\to \tt$ respectively.  We will use `ISR' to refer to this model.

A more sophisticated approximation would also include final-state radiation
by allowing the $b$ quarks to radiate,
assuming that they were linked by a second color string, as for example
for the process  $q \bar q \to Z^0 \to \bb$.
   In this case we would find\footnote{This
would in fact be the correct result for the process $\qq (gg)
\to H \to \tt \to \bb\ww$ in the limit $\Gamma_t \to \infty$ \cite{KOSHAD}.}
\begin{equation}
\cF\;  = \;2 C_F (2N)  \; \akikii \;
+ \; C_F \;[ 2\apipii - \apipi- \apiipii ] \; .
\label{isrfsronly}
\end{equation}
We will refer to this model as `ISR/FSR'.
The ISR and ISR/FSR models are easily implemented and are characteristic of
what appears in parton-shower Monte Carlos.

A third possible model corresponds to the naive expectation that the top quarks
do not radiate because of their short lifetime.  (In our case, where the
relevant gluon energies are larger than $\gt$, the top quarks {\it do}
radiate; see \cite{KOS} for a discussion.)
This model corresponds to taking
$\gt\to\infty$ (and hence $\chi_i, \chi_{12}\to 1$)
in Eq.~(\ref{chidef}) and gives
\begin{equation}
\cF\;  = c_1 \akikii + c_2 \akipi +
c_3 \akipii + c_4 \akiipi
+ c_5 \akiipii + c_6 \apipii + c_7 \apipi + c_8 \apiipii\; .
\label{bb}
\end{equation}
This is the radiation pattern we would see if the $b$ and $\bbar$ were
produced directly, and so we will refer to this model as `BB'.

In what follows, we will compare the complete distribution given in
Eq.~(\ref{general}) with those from the ISR, ISR/FSR, and BB models,
corresponding to Eqs.~(\ref{isronly}), (\ref{isrfsronly}), and (\ref{bb}).

\section{Numerical results and discussion}

In the previous section we have seen how the radiation pattern depends
on the orientation of the final-state $t$ and $b$ quarks.
This obviously varies from event to event, and in our calculations
we must therefore integrate over the full phase space
 for $\tt$ production and decay
and weight by the appropriate parton distributions. This will then yield
the correct radiation pattern in the laboratory frame.
In particular, we will see important differences between the models
discussed at the end of the previous section. To understand how and why
these differences arise, it is  helpful to first work in the
parton subprocess center-of-mass, fixing the $t$ and $b$ momenta
in `typical' configurations, as was done
in Refs.~\cite{KOS,KOSHAD}.

In order to assess the typical final-state configurations, we first
generate a sample of events corresponding to
$\qq, gg \to \tt \to \bb \ww \to \bb ll\nu\nu$
with $m_t = 174\ \GeV/c^2$ in $\pp$ collisions
at $1.8\ \TeV$.
Figure~{\kinem} shows the resulting distributions in $b(\bar b)$
transverse momentum
and  pseudorapidity,  and in the $b-\bar b$
azimuthal angle difference. The majority of $b$ quarks are produced
centrally ({\it i.e.},
within $ |\eta_b| < 1$) and with $p_T^b \sim 30 - 100 \ \GeV$.
The distribution in azimuthal angle difference shows a slight preference
for back-to-back production in the transverse plane.
We note for future reference that the centrality of the $b$'s is due in large
part to the fact that
they receive a boost in the direction of their parent top quarks, which tend
to be produced centrally; this also accounts for their slight back-to-back
preference.
We note also that for $m_t = 174 \ \GeV/c^2$, the cross section is dominated
by the $\qq \to \tt$ process, which is an order of magnitude larger
than the $gg \to \tt$ contribution.

\subsection{Features of the gluon distribution in the $\tt$ center of mass}

To investigate the radiation pattern of soft gluon emission, we choose
a configuration with the following properties. For the moment
we work in the $\tt$
center-of-mass frame, with subprocess energy $\sqrt{\hat{s}} = 3 m_t$.
The $t$ and $\bar t$ are produced centrally ($\eta_t = \eta_{\bar t} = 0$)
and the $b$ quarks have $\eta_b = -\eta_{\bar b}$ and $\Delta\phi_{\bb} =
180\degree$, with the direction of the $b$ defining $\phi = 0\degree$.
For purposes of illustration, we will consider the particular case
$\eta_b =  1$, which corresponds to $p_T^b = p_T^{\bar b} = 57\; \GeV/c$.
The distribution of gluon radiation in the $\eta - \phi$ plane
is then obtained from Eq.~(\ref{softsigma}):
\begin{equation}
{d N\over  d\eta_g \> d\phi_g}\ \equiv \
{1\over \cosh^2\eta_g}\ {1\over d\sigma_0}
\ {d\sigma\over d E_g\> d\cos\theta_g \> d\phi_g}\ = \
{\alpha_s\over 4 \pi^2} \ { E_g \; \cF \over \cosh^2\eta_g } \; .
\label{legodist}
\end{equation}
At this stage we are only concerned with the general features
of the radiation pattern, and so we (somewhat arbitrarily) set
$E_g = 10\ \GeV$ and $\alpha_s = 0.1$.

Figure~{\threed} shows the distribution of Eq.~(\ref{legodist}), for the choice
$\eta_b = 1$. The two most obvious features are: (i)  the strong enhancement
of the radiation near the $b$ and $\bar b$ directions, and (ii) the
approximately constant `pedestal' distribution away from the $b$ quark
directions and in particular at large positive and negative $\eta_g$.
Closer inspection reveals another important feature -- {\it the radiation
has minima close  to the original $t$-quark directions}, in this
case at $\eta_g \approx 0$. These represent  the `dead cones' of the $t$ and
$\bar t$ quarks \cite{KOS}.
In order to study the various individual
contributions to the distribution, and to
compare different approximations, it is more useful to consider
particular slices through the two-dimensional plot.
In Fig.~{\slicesa}  we show the same distribution as a function of $\eta_g$ for
the particular values $\phi_g
= 0\degree,\;45\degree,\; 90\degree$. The curves correspond to the
decomposition of Eq.~(\ref{general}): the long-dashed line is the
\lq production' contribution, the short-dashed line is the \lq decay'
contribution, the dotted line is the interference contribution,
and the solid line is the total. For $\phi_g  (= \phi_b = \phi_t) =
0\degree$, we clearly see the dead cone of the $b$ quark (in the decay
contribution) and  the much broader dead cone of the $t$
quark (in the production
contribution). The interference contribution is small. Away from the
$b$-quark direction, the production contribution dominates.
At $\phi_g = 45\degree$ the effects of the final-state quarks
on the radiation pattern are much reduced, and at $\phi_g = 90\degree$
-- which is exactly half-way between the $b$ and $\bar{b}$ quarks --  the
production contribution dominates and the distribution is
approximately uniform.

Figure~{\slicesb} shows the total contribution of Fig.~{\slicesa}
(solid line)
compared to the distributions corresponding to
 the ISR (dashed line) and ISR/FSR
(dotted line) models.  The latter is approximately equal to the
correct distribution close to the $b$-quark direction and at large
forward and backward rapidities. However, it lacks the $t$-quark
dead-cone effect, and overestimates the radiation in the region
between the $b$ and $\bar b$ quarks. The ISR model gives a
constant contribution, and is in strong disagreement with the
correct distribution  for $\vert \eta_g \vert \lapp 2$.

Although Figs.~{\slicesa} and {\slicesb} are useful are illustrating particular
features of the various models, they correspond to idealized
situations where the momenta of the
$t$ and $b$ quarks are fixed. In the next section,
we present more realistic  distributions corresponding to the fully integrated
$t$- and $b$-quark cross sections.

\subsection{Full gluon distributions in $\tt$ events at the Tevatron}

Here we obtain the full distribution of gluons expected in $\tt$ events at the
Tevatron, for the process $\qq\to\tt\to\bb\ww$,
with gluons generated according to Eq.~(\ref{softsigma}).  We use MRS(H)
parton distributions \cite{MRSH}.  We neglect the $gg\to\tt$ process, whose
contribution to the cross section is an order of magnitude smaller than that
for the $\qq$ initial state.  This simplifies  the
interpretation of the results
because the color structure of the two processes is different \cite{KOSHAD}.

Our calculations are performed entirely at the parton level, and
kinematic cuts are
kept to a minimum.  We require that the $b$'s are produced centrally
and that the gluons fall within some typical detector pseudorapidity range.
We also require that the gluons have sufficient transverse momentum to be
detectable as soft
jets but not so much as to invalidate the soft approximation.
Finally, we require some angular separation between the $b$'s and gluons so
that
their respective jets are distinguishable.
These requirements are implemented through the following kinematic cuts:
\begin{eqnarray}
|\eta_{b}|,|\eta_{\bbar}| \> & \leq & \> 1.5 \; ,\nonumber \\
|\eta_g| \> & \leq & \> 3.5 \; ,\nonumber \\
10\ \GeV/c \leq \> & p_T^g & \> \leq  25\ \GeV/c \; ,\nonumber \\
E_g \> & \leq & \> 100\ \GeV \; ,\nonumber \\
\Delta R_{bg},\Delta R_{\bbar g} \> & \geq & \> 0.5 \; .
\label{cuts}
\end{eqnarray}
Note that the $\eta_g$ cut eliminates the collinear singularities associated
with initial-state radiation.
With the exception of the cuts on $\Delta R$ and to some extent $\eta_g$,
our results are not particularly sensitive to the exact values used.

Energy and  transverse momentum distributions
of the resulting gluons are shown as solid lines in Figure {\ptet}(a) and (b).
There are no surprises; the figures display the
expected fall-off with increasing energy and $p_T$,  and
reflect the lower and upper cutoffs.
Also shown in Fig.~{\ptet} are the individual contributions
according to
the decomposition in Eq.~(\ref{general}), with contributions from production
and decay appearing as dotted and dashed lines, respectively.  The interference
contribution is negligible for the energies considered here and is not shown
in this plot or any that follow.
We see that the energy distribution from the production piece extends to
higher energies than that from decay, and that
 the decay $p_T$ distribution is relatively flatter than that from production.

\subsubsection{Angular distribution and top mass reconstruction}

The difference between the distributions of gluons produced in the production
and decay stages is displayed more dramatically in the gluon pseudorapidity
distribution, Figure {\etapd}.
We are most interested in angular distributions in any case
because we want to know where
the soft jets are likely to appear in detectors.
Figure {\etapd} shows the net $\eta_g$ distribution (solid
line) along with the production--decay decompostion as in the previous figure.
  The decay contribution is peaked in the center and
falls off quickly, whereas the production piece  has a central dip and
peaks in the forward direction.
These results are easily understood when we consider the sources of the
gluons; {\it cf.} Eq.~(\ref{general}).
Gluons from the decay contribution
come from radiation off the top and bottom quark lines.  Radiation is largest
near the quarks' direction of motion, and the quarks are, for the most part,
produced centrally.  Furthermore, we note that the string effect in
the $t$--$b$ and $\tbar$--$\bbar$
antenna pieces (the $\aqipi$ and $\aqiipii$ terms in $\cF{\mbox{\tiny DEC}}$)
tends to enhance central radiation compared to that in the forward regions
when the quarks are produced centrally.
Similarly, the production piece arises from radiation off initial-state quarks
and the $t$ and $\tbar$, and we get enhancement in the forward direction.
The net effect is a total distribution that is slightly
peaked in the center and falls off sharply above $\eta_g=2$ or so.

Let us now compare these distributions with what we obtain using the simpler
models described above.  It seems plausible that the ISR/FSR model
(initial-state and final-state radiation only, assuming color singlet
initial and final states) would give a reasonable approximation to the
correct distribution, with ISR corresponding to the production piece and
FSR corresponding to decay.
Figure {\models} shows net $\eta_g$ distributions for the various models,
with the solid line again indicating the correct
distribution according to Eq.~(\ref{general}).
We see that the ISR model gives the flat rapidity distribution
characteristic of initial-state radiation.  Comparison with the production
contribution in  Fig.~{\etapd}
(note the difference in vertical scales) shows that the ISR model significantly
overestimates the amount of radiation in the central region.  Adding
to that radiation off the final $b$'s to obtain the ISR/FSR
distribution (dashed lines)
leads to an even greater overestimate.
Hence
 the ISR/FSR model
is not a particularly good approximation to the full prediction: the overall
normalization is too large and the total shape is wrong, being  too
strongly peaked in the center.

The other simple model --- the BB model, in which the
radiation is treated as in the direct $\bb$ production process $\qq\to\bb$
--- {\it does} in fact  reproduce the net $\eta_g$ distribution reasonably
well;
see the dot-dashed line in Fig.~{\models}.  Despite the fact that, for the
gluon energies of interest here, the top width $\gt$ can  be considered
to be small, the $\gt\to\infty$ limit seems to yield a good
approximation to the full prediction,
{\it i.e.}, the  results seem to be independent of the $\chi_i$ and $\chi_{12}$
factors in Eq.~(\ref{general}).
The reason for this is related to the fact that radiation close to  the
$b$ quarks  is universal, independent of the process (see Ref.~\cite{KOSHAD}
for a discussion).
But this result is deceptive for a number of reasons.  First, depending as it
does on radiation near the $b$'s, it  is sensitive to our choice of
$\Delta R$ cut; if the gluons are restricted to be farther away from the $b$'s,
the agreement deteriorates.  More important, although the $\eta$ distribution
seems to be well reproduced, the distribution in azimuthal angle
 is not, as we shall
see below.  Finally, because radiation off the $t$ quarks is ignored, there is
no way to decompose the results into production and decay contributions.

We now address the question, ``What should be done with extra soft jets in
attempts to reconstruct the top quark four-momentum?"
As stated above, the decomposition of the gluon distributions into
contributions from gluons radiated in top production and those
radiated in decay provides guidance.
Whether additional gluons
in $\tt$ events should be combined with the $b$ and $W$ depends on
what stage they are associated with:  production gluons should not be combined,
and decay gluons should.  In principle there is also a contribution from
interference terms which cannot be definitely associate with either stage,
but in practice, for gluons sufficiently energetic to be identified as
soft jets, these interference terms are negligible.

With this in mind, we return to the decomposition of the gluon
pseudorapidity distribution shown in Fig.~{\etapd}.
We saw  that gluons in the forward region are much more likely to have
come from the production than the decay stage.  Hence most forward
gluons do not contribute to the top momentum, and based on these
parton--level results it seems safe to conclude that soft jets with
$|\eta_g|>1.5$
can be ignored without introducing large systematic errors into the
mass measurement.  (This assumes, of course, that such jets are correctly
identified.)
Unfortunately the situation is less clear-cut in the central region.
Despite the central suppression of production-stage gluons, Fig.~{\etapd}
shows that gluons in the central region are almost as likely to have
come from production as from decay; in fact, with a larger $\Delta R$ cut,
central gluons would be dominated by production.
Note that the situation is even worse for the simpler models,
since the ISR $\eta_g$ distribution is flat, and the
BB model allows for no decomposition at all.

In an attempt to identify a quantity that shows a clear excess in
the decay contribution, we consider proximity to the $b$ or $\bbar$ quark.
It is certainly plausible to expect that soft jets near the
$b$'s are likely to have come from decay--stage gluons.
Figure {\cosbg}(a) shows the gluon distribution as a
function of the cosine of the angle between the $b$ and $g$,
with the usual decomposition into production and decay contributions
(dots and dashes, respectively), and total (solid).  The production piece is
flat, as expected since there is no dependence on the $b$ direction in
$\cF_{\mbox{\tiny PROD}}$,
and the decay piece increases as the gluon approaches the $b$.
At first sight this looks promising -- there is an excess in the decay piece
near the $b$.  However, this result should not be taken quite at face value
because the amount of this excess is extremely
sensitive to the cuts, especially that on  $\Delta R$ between the gluon
and $b$.  The excess can be reduced or even eliminated by reducing the gluon
$p_T$ range or increasing the $\Delta R$ cut.  This would be compounded in
practice by fragmentation of the partons into hadrons.

We can improve the situation by taking advantage of the difference
in the production and decay  pseudorapidity distributions and
making a tighter $\eta_g$ cut to eliminate
much of the production contribution while retaining almost all of that from
decay.  Figure {\cosbg}(b) shows that requiring $|\eta_g|\leq 1.5$ does
indeed enhance the latter --
 the excess in the region near the $b$ quark is now more pronounced.
Unfortunately, the same caveats about cuts and fragmentation apply here, and
we cannot safely quantify this excess or draw any firm conclusions.

It is clear from what we have seen above that assignments of soft jets to the
production or decay stages cannot be made unambiguously on an event-by-event
basis.  However, we can minimize errors associated with soft jet ambiguities
if we know what to expect from the correct gluon distribution and its
decomposition and include it in our simulations.
Eventually we may hope to find additional discriminators that
improve the chances of making correct assignments,
leading ultimately to a useful prescription for dealing with soft jets.
The results shown here represent a first step in that direction.
We do not perform mass reconstruction simulations here, since we do not
use exact kinematics in the soft gluon approximation.  In future work
we intend to dispense with the soft approximation and perform the exact
calculation, which will allow us to investigate more specifically
issues associated with mass reconstruction.


\subsubsection{Forward--backward asymmetry and color structure}

We close this section by noting that the color structure of the process
 process $\qq\to\tt\to\bb\ww$ can
give rise to distinctive effects in the gluon radiation pattern that
 can be explored experimentally, at least in
principle.  In particular, a forward--backward asymmetry appears  in the
distribution of gluons for an appropriately chosen
class of final-state configuations.
While not directly relevant to top mass measurement, this asymmetry
helps to illuminate further some of the physics involved, for example
the fact that the top quarks themselves can radiate gluons.

This asymmetry arises as a consequence of the string effects in
the various antennae, that is, the enhancement of radiation
between two quarks connected by a color string \cite{MW} (see also
\cite{BOOK}).  For example, in $\qq\to\tt$
the $\widehat{qt}$ antenna produces more radiation in the
region between the $t$ and $q$ than, say, between the $t$ and $\qbar$.
This gives rise to a forward--backward asymmetry in the gluon
radiation.
This asymmetry is canceled in the net distribution by the
$\widehat{\qbar\tbar}$
antenna, of course.  However, with suitably chosen cuts we
can enhance one contribution (from radiation off the $t$, say)
while suppressing the other, thereby recovering the asymmetry.
We do this by taking advantage of the fact that there is more
radiation from a quark in the regions nearby than those far away.

Guided also by the fact that the
direction of the  $b$ quarks' momenta tend not to be too different from
those of their parent $t$'s,
we add to the event selection criteria of Eq.~(\ref{cuts}) the
requirement that the
separation in azimuthal angle between the $b$ and $\bbar$ be greater than
$135\degree$.  This tends to select events in which the parent
$t$ and $\tbar$
have similar separation.   We then preferentially select gluons associated
with the $t$ (as opposed to the $\tbar$) by requiring that they lie
within $90\degree$ in azimuth of the $b$ quark.  Figure~{\asym}(a) shows
the resulting distribution as a solid line.  A clear excess is visible in
the forward direction, which is the direction of the initial quark's
momentum.  An equal and opposite asymmetry is obtained by considering gluons
within $90\degree$ of azimuth of the $\bbar$.

That the source of this asymmetry is indeed the terms involving the initial
quark and the $t$ and $\tbar$ ({\it cf.} Eq.~(\ref{general})) can be seen from
the decomposition shown in Fig.~{\asym}(a), where, as in earlier figures, the
production contribution is shown as the dotted line and the decay
contribution is the dashed line.  We see that the entire asymmetry comes
from the production piece.  The decay piece involves only the top and bottom
momenta and there is no correlation with the initial quarks.

Finally, the asymmetry distribution reveals major discrepancies between the
correct distribution and all three of the simpler models, as shown in
Figure {\asym}(b).
In the ISR (dotted line) and ISR/FSR (dashed line) models, in which there
is no connection between radiation from  the initial and final states,
there can be no forward--backward asymmetry.  In contrast, the
BB model has a more pronounced asymmetry even than the correct
distribution because, without radiation
from the top quarks, the antennae connecting initial- and final-state quarks
are enhanced.  This difference between BB and the correct result
shows that the BB model does not contain the correct azimuthal dependence,
despite the fact that it appears to reproduce the correct total $\eta_g$
distribution in Fig.~{\models}.

\section{Conclusions}

The pattern of soft gluon radiation in $t\bar t$ production in hadronic
collisions has a very rich structure. The initial-state partons, the
$t$ and $b$ quarks can all radiate, and the gluon distribution in
any part of phase space is a combination of what can be termed
`production', `decay' and `interference' contributions.
Apart from providing some interesting tests of the color structure
of the events, there are important implications for top mass
measurements from the invariant mass of the decay products.
In this study we have attempted to illustrate the general features
of the radiation pattern. We have quantified the idea that soft
jets close to the $b$ quarks can be associated with gluon emission in the
top decay process, while radiation at large rapidities can be associated
with radiation at the $t \bar t$  production stage. We have compared
the full QCD prediction with those of simpler models, which are
characteristic of how radiation may be implemented in Monte Carlo
event generators. It is not difficult to find regions of phase space
where the models differ significantly from the full prediction.
Our conclusion is that the question of how to handle additional
soft jets in $t \bar t$ events when reconstructing the top momentum
is far from trivial.
In a future study we intend to go beyond the soft gluon approximation
to study the invariant mass distributions themselves.

\bigskip
\medskip
\noindent{\Large\bf Acknowledgements}
\bigskip

\noindent One of us (WJS) is grateful to the UK Science and Engineering
Research Council for a Senior Fellowship, and to the University
of Rochester for hospitality during the early stages of this work.
Useful discussions with Valery Khoze are acknowledged.
This work was supported in part by the U.S.\ Department of Energy,
under grant DE-FG02-91ER40685.
\goodbreak


\vskip 1truecm

\section*{Figure Captions}
\begin{itemize}
\item [{[\kinem]}]
Distributions in (a) the $b$-quark transverse momentum, (b) the
$b$-quark pseudorapidity, and (c) the azimuthal angle difference between the
$b$ and $\bar b$ quarks, in $\tt$ production, via the subprocesses
$\qq, gg \to \tt \to \bb \ww $, in $\pp$ collisions at $\sqrt{s} =
1.8\ \TeV$.

\item [{[\threed]}]
Soft gluon radiation pattern in the $\eta - \phi $ plane, for
 the $\qq\to \tt$ process in the $\tt$ center-of-mass
frame. The $t$, $\bar t$,  and $b$, $\bar b$ quarks are at
$(\eta, \phi ) = (0,0\degree ), \; (0,180\degree )$
and $(\eta, \phi ) = (1,0\degree ), \; (-1,180\degree )$ respectively.

\item [{[\slicesa]}]
The soft gluon distribution of the previous figure
for fixed values of $\phi_g$.  The curves correspond to the
decomposition of Eq.~(\ref{general}): the long-dashed line is the
\lq production' contribution, the short-dashed line is the \lq decay'
contribution, the dotted line is the interference contribution,
and the solid line is the total.

\item [{[\slicesb]}]
The total soft gluon distribution of Fig.~\slicesa\ (solid line)
together with  the distributions corresponding to
 the ISR (dashed line) and ISR/FSR (dotted line) models.

\item [{[\ptet]}]
Distributions in (a) the gluon energy and (b) the gluon transverse momentum,
in $\tt$ production, via the subprocess
$\qq\to \tt \to \bb \ww $, in $\pp$ collisions at $\sqrt{s} =
1.8\ \TeV$.  Contributions from production (dotted lines) and decay (dashed
lines) are shown along with their totals (solid lines).  The cuts are
listed in  Eq.~(\ref{cuts}).

\item [{[\etapd]}]
Gluon pseudorapidity distributions in $\tt$ production, via the subprocess
$\qq\to \tt \to \bb \ww $, in $\pp$ collisions at $\sqrt{s} =
1.8\ \TeV$.  The net distribution is shown as a solid line; contributions from
production (dotted lines) and decay (dashed lines) are also shown.
The cuts are listed in Eq.~(\ref{cuts}).

\item [{[\models]}]
Gluon pseudorapidity distributions in $\tt$ production, for the models
discussed in Section 2:  full  distribution (solid line), ISR model (dotted
line), ISR/FSR model (dashed line), and BB model (dash-dotted line).
The cuts are listed in  Eq.~(\ref{cuts}).

\item[{[\cosbg]}]
Distribution in the cosine of the angle between the gluon and the $b$ quark,
 (a) with cuts as in Eq.~(\ref{cuts}) and
 (b) with the additional cut $|\eta_g|\leq 1.5$.

\item [{[\asym]}]
Forward--backward asymmetry in gluon pseudorapidity distributions in
$\tt$ production.  The cuts are as in Eq.~(\ref{cuts}) with the additional
requirements $\Delta\phi_{\bb}>135\degree$ and $\Delta\phi_{bg}<90\degree$.
The curves correspond to the
(a) total (solid), production (dots) and decay (dashes) distributions, and
to the  distributions for the
(b)  full QCD  (solid),  ISR (dots), ISR/FSR (dashes)
and BB (dash-dots) models.

\end{itemize}

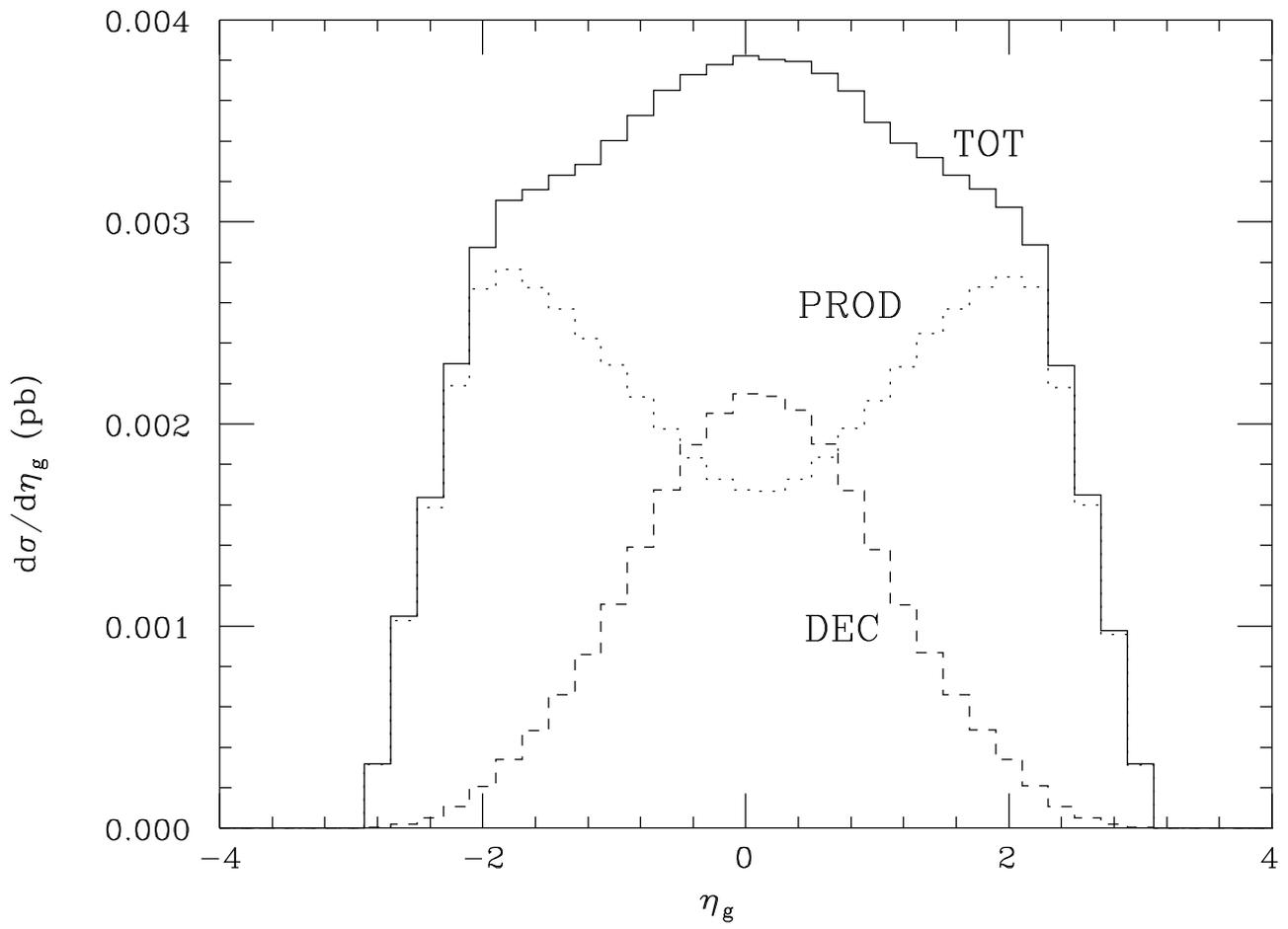
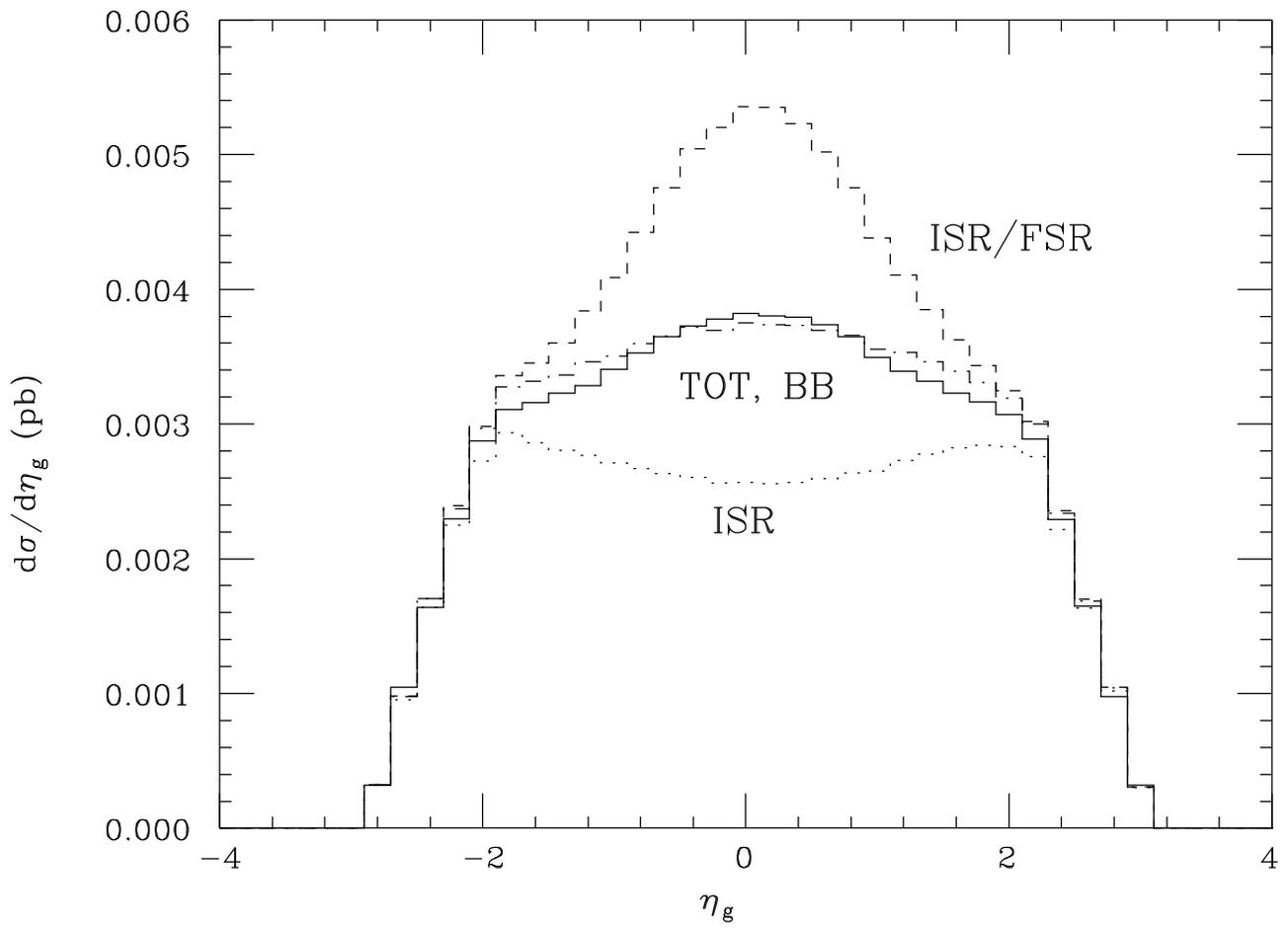
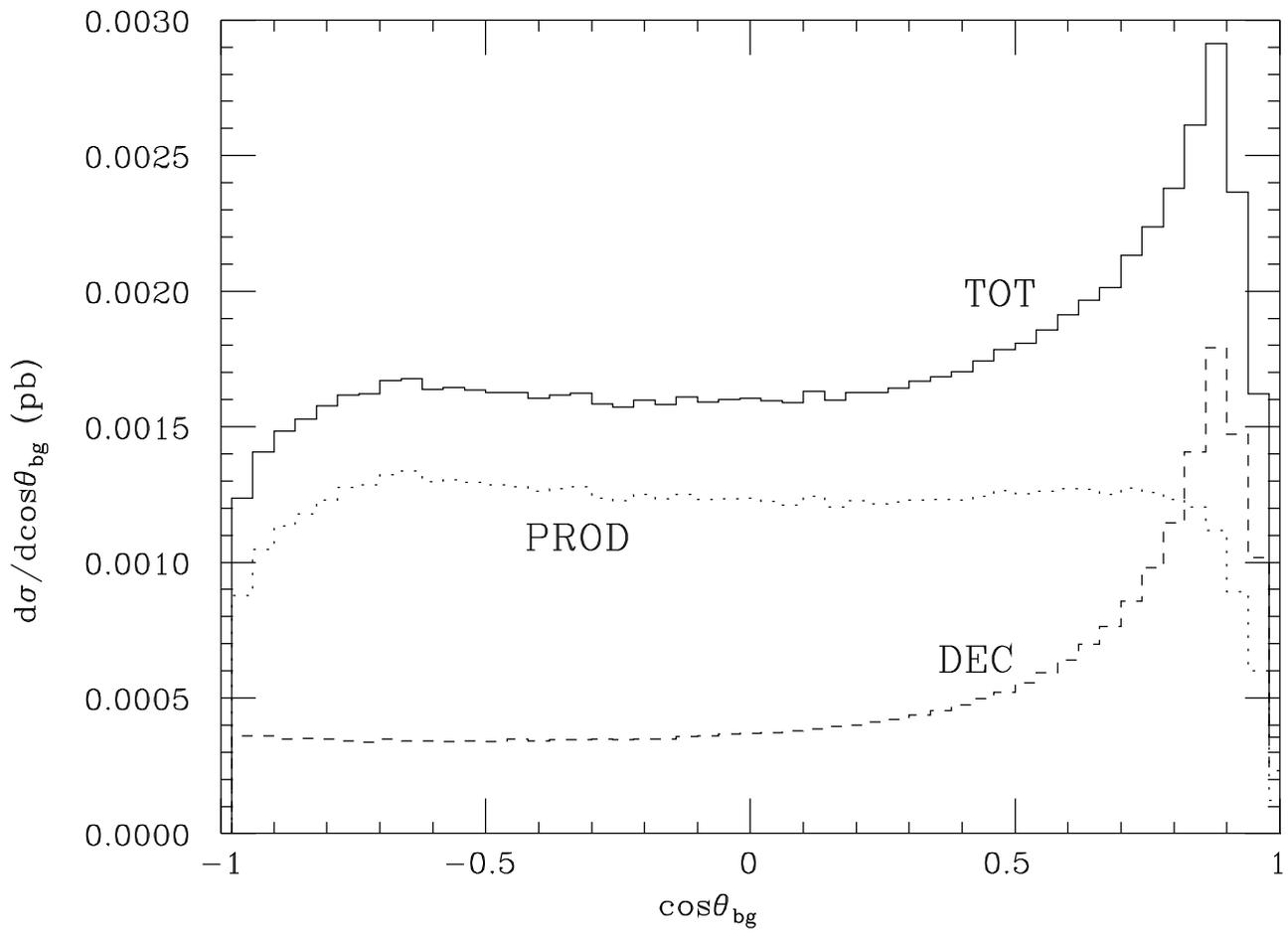
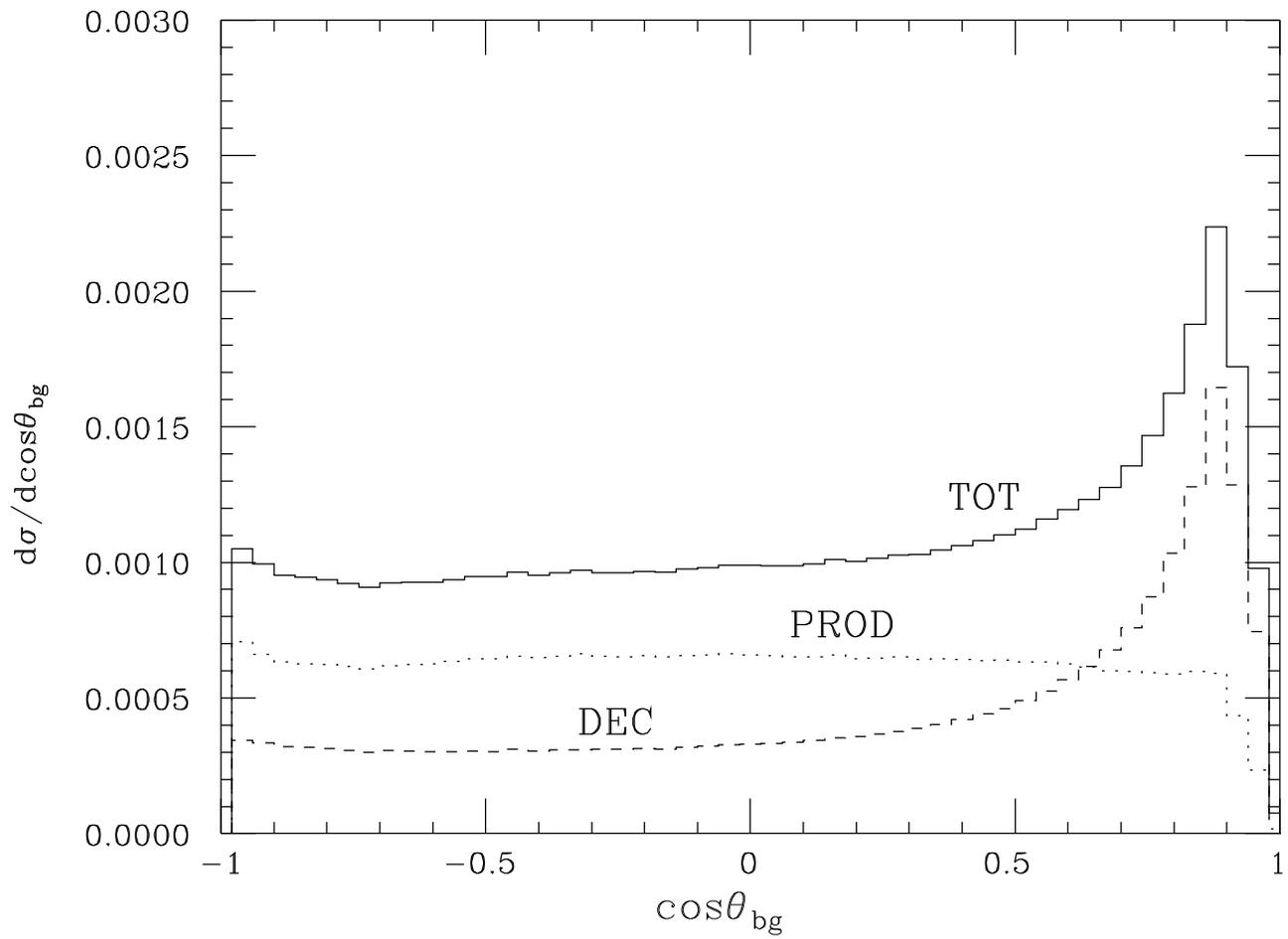
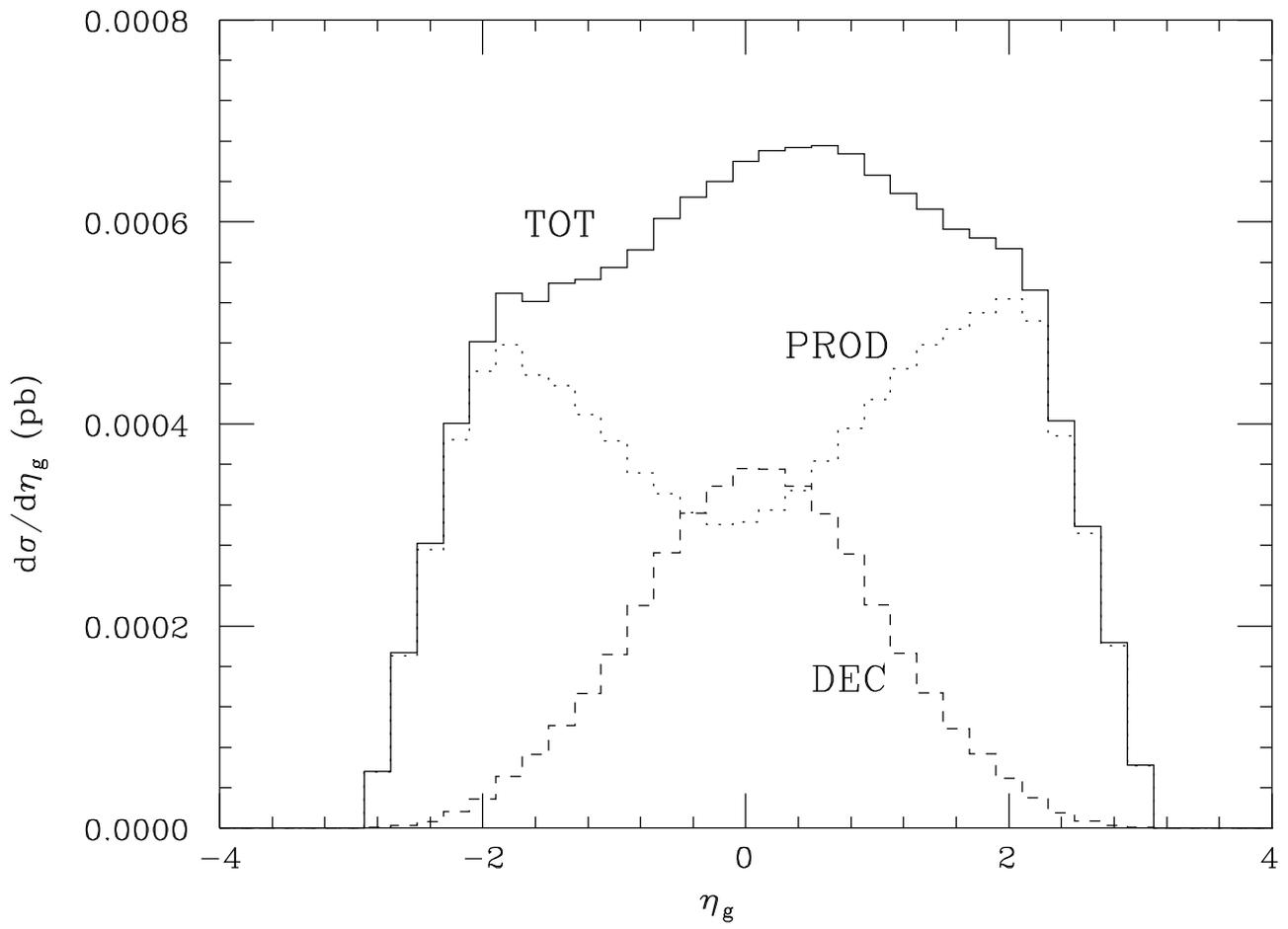
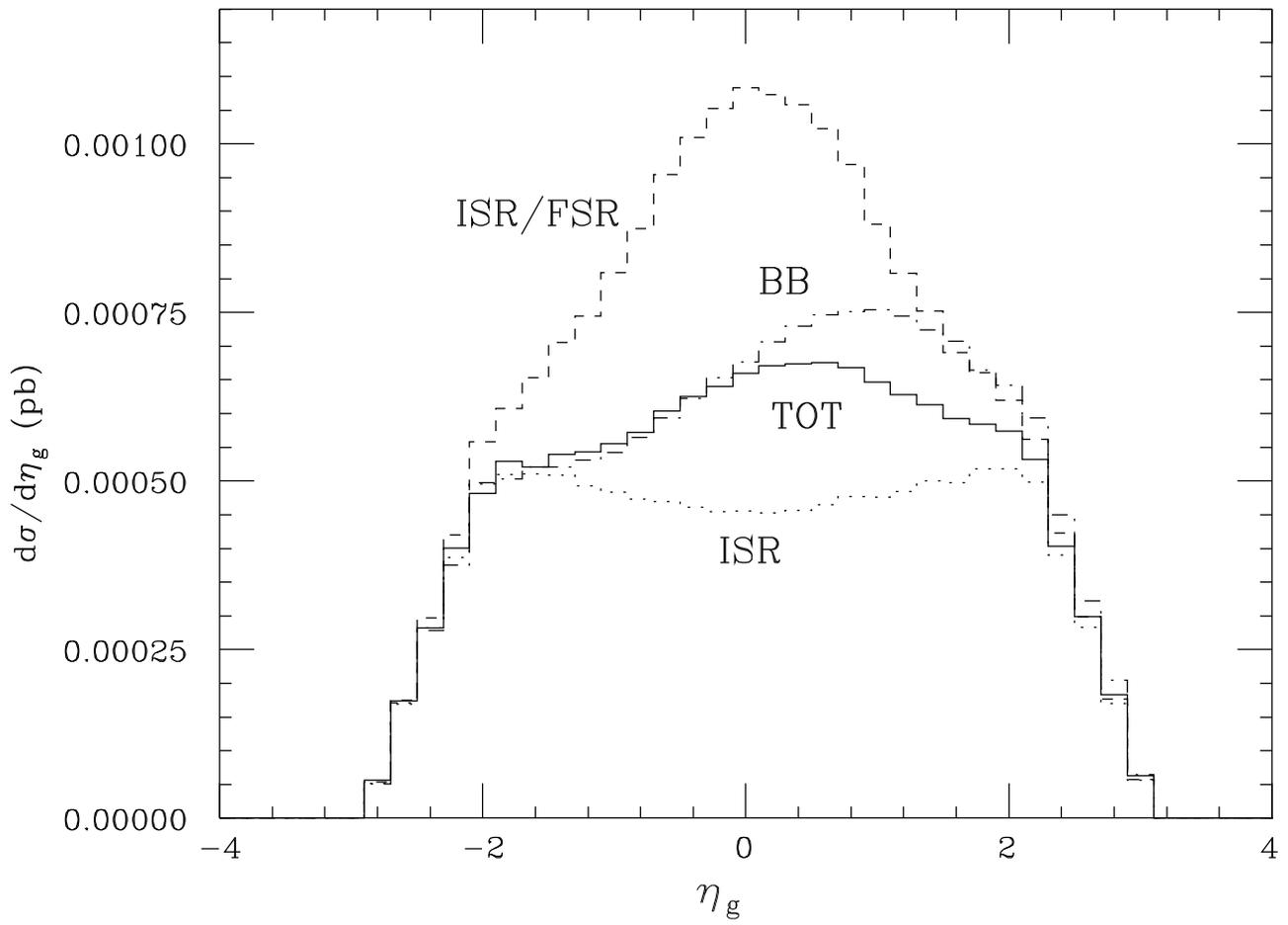
\end{document}